\documentclass[aps,preprint,amsmath,amssymb,superscriptaddress,titlepage]{revtex4-2}
\usepackage{graphicx}
\usepackage{hyperref}
\usepackage{bm}
\usepackage{color}
\hypersetup{colorlinks=true}
\hypersetup{
    breaklinks = true,
    colorlinks = true,
    citecolor = {blue},
    urlcolor = {blue},
    linkcolor = {blue}
}

\begin{document}



\title{Photon-polarization-resolved linear Breit-Wheeler pair production in a laser-plasma system}

\author{Huai-Hang Song}\email{huaihangsong@sjtu.edu.cn}
\affiliation{Key Laboratory for Laser Plasmas (MOE) and School of Physics and Astronomy, Shanghai Jiao Tong University, Shanghai 200240, China}
\affiliation{Collaborative Innovation Center of IFSA, Shanghai Jiao Tong University, Shanghai 200240, China}

\author{Zheng-Ming Sheng}
\affiliation{Key Laboratory for Laser Plasmas (MOE) and School of Physics and Astronomy, Shanghai Jiao Tong University, Shanghai 200240, China}
\affiliation{Collaborative Innovation Center of IFSA, Shanghai Jiao Tong University, Shanghai 200240, China}
\affiliation{Tsung-Dao Lee Institute, Shanghai Jiao Tong University, Shanghai 201210, China}

\date{\today}

\begin{abstract}

The linear Breit-Wheeler (LBW) process, mediated by photon-photon collisions, can emerge as the dominant pair production mechanism in the ultraintense laser-plasma interaction for laser intensities below $10^{23}~\rm W/cm^2$. Here, we explore the role of photon polarization in LBW pair production for a 10 PW-class, linearly polarized laser interacting with a solid-density plasma. The motivation for this investigation lies in two main aspects: photons emitted via nonlinear Compton scattering are inherently linearly polarized, and the LBW process exhibits a distinct sensitivity to photon polarization. By leveraging particle-in-cell simulations that self-consistently incorporate photon-polarization-resolved LBW pair production, our results reveal that photon polarization leads to a 5\% to 10\% reduction in the total LBW positron yield. This suppression arises because the polarization directions of the colliding photons are primarily parallel to each other, resulting in a diminished LBW cross section compared to the unpolarized case. The influence of photon polarization weakens as the laser intensity increases.

\end{abstract}


\maketitle


\section{Introduction}

The linear Breit-Wheeler (LBW) process describes electron-positron ($e^\pm$) pair production through the collision of two high-energy photons \cite{breit1934pr}. The condition for LBW pair production to occur is that the energy of two colliding photons in the center-of-momentum (CM) frame---where the net momentum of two photons is zero---must exceed the rest mass energy of the electron $m_e c^2 \approx 0.511$ MeV, where $m_e$ is the electron mass and $c$ is the speed of light in vacuum. Besides, the involved photons must be sufficiently brilliant to produce a significant number of $e^\pm$ pairs, as the maximum of the LBW cross section is just on the order of $10^{-25}$ cm$^2$. The long-standing absence of high-energy and brilliant photon sources has hindered the experimental observation of LBW pair production directly using real photons, although its another form using virtual photons has been experimentally confirmed \cite{adam2021prl}.

The development of high-power laser facilities in PW and 10 PW classes \cite{li2022hplse,tanaka2020mre,burdonov2021mre} has opened unprecedented opportunities to investigate strong-field quantum electrodynamics (QED). The interaction strength in the strong-field QED regime is primarily characterized by a parameter $\chi_e = \frac{e\hbar}{m_e^3 c^4} |F_{\mu\nu} p^{\nu}|$, where $F_{\mu\nu}$ is the electromagnetic field tensor, $p^{\nu}$ is the electron four-momentum, $e$ is the elementary charge, and $\hbar$ is the reduced Planck constant. The energetic electron can lose a significant portion of its energy by emitting individual $\gamma$ photons via nonlinear Compton scattering \cite{cole2018prx,poder2018prx,blackburn2020rmpp}. The plasma, acting as an efficient medium, enables significant conversion of laser energy into $\gamma$ photons \cite{nerush2011prl,ridgers2012prl,brady2012prl,ji2014prl,stark2016prl,benedetti2018np,wang2018pnas,vranic2019pop}. The use of laser-plasma-driven photon sources to probe LBW signals has garnered considerable attention \cite{pike2014np,ribeyre2016pre,yu2019prl,wang2020prl,he2021cp,song2024pre}. Recently, a conventional laser-solid setup has been proposed to explicitly distinguish LBW pairs from other underlying mechanisms \cite{song2025arxiv}, which greatly simplifies experimental challenges. In addition to validating this fundamental QED process, generating quasimonoenergetic positrons seeded by the LBW process also provides new insights for positron applications \cite{sugimoto2023prl,song2025arxiv}. Previous studies on pair production in ultraintense laser-plasma systems have primarily focused on the nonlinear Breit-Wheeler (NBW) process \cite{reiss1962jmp}, which requires significantly higher laser intensities \cite{nerush2011prl,ridgers2012prl}. For laser intensities below $10^{23}~\rm W/cm^2$ that are currently achievable \cite{yoon2021optica}, the LBW process becomes dominant and thus warrants particular attention.

To more accurately predict LBW pair production in the laser-plasma interaction, it is essential to consider photon polarization for two reasons. First, photons emitted via nonlinear Compton scattering in the strong-field QED regime exhibit a high degree of linear polarization, with polarization degrees surpassing 50\% \cite{li2020prl,xue2020mre,song2021njp,gong2022prr}. Second, the LBW cross section depends on photon polarization, which affects the positron yield and positron emission direction \cite{breit1934pr,ginzburg1984nimpr,zhao2022prd,zhao2023prd}. In astrophysics, for example, neglecting photon polarization of synchrotron-radiated photons can lead to an overestimated $\gamma\gamma$ opacity by approximately 10\% \cite{bottcher2014apj}. In laser-plasma systems, this issue remains unresolved as it has only recently become possible to self-consistently simulate polarized nonlinear Compton scattering \cite{xue2020mre,song2021njp,gong2022prr} and unpolarized LBW pair production \cite{song2024pre,song2025arxiv} by modern particle-in-cell (PIC) simulations.

In this paper, we introduce how to implement the photon-polarization-resolved LBW process into a PIC code. Based upon this, the photon polarization effect on LBW pair production is investigated in the interaction of a 10 PW-class, linearly polarized laser with a solid-density plasma. We find that photon polarization reduces the total LBW positron yield by 5\%–10\% compared to the unpolarized case. This reduction arises because photons emitted via nonlinear Compton scattering are primarily linearly polarized within the laser polarization plane. After transforming to the CM frame, the polarization directions of the colliding photons are mostly parallel, leading to a smaller LBW cross section compared with the unpolarized one. The suppression due to photon polarization is angle-dependent, with a stronger effect on positrons emitted in directions perpendicular to the laser polarization plane. As the laser intensity increases, the photon energy rises while photon polarization decreases, both of which results in a diminished effect of photon polarization on LBW pairs.

The rest of this article is organized as follows. Section \ref{algorithm} introduces the theory and numerical method related to photon polarization in nonlinear Compton scattering and LBW pair production. In Sec.~\ref{simulation}, we present our PIC setup and simulation results for a solid-density plasma irradiated by a linearly polarized laser, with a particular focus on polarization of emitted photons and its effect on LBW positrons. Section \ref{conclusion} provides a brief summary.

\section{Two polarized QED processes}
\label{algorithm}

\subsection{Photon-polarization-resolved nonlinear Compton scattering}
\label{compton}

The photon-polarization-resolved rate of nonlinear Compton scattering under the locally constant field approximation is given by \cite{baier1998qed}
\begin{eqnarray}\label{eq1}
\frac{d^2W_{\rm rad}}{dudt}=\frac{\alpha m_e^2c^4}{\sqrt{3}\pi \hbar \varepsilon_e}\bigg[\frac{u^2-2u+2}{1-u}K_{2/3}(\kappa)-{\rm Int}K_{1/3}(\kappa)+\xi_3K_{2/3}(\kappa)\bigg],
\end{eqnarray}
where $\alpha\approx 1/137$ is the fine structure constant, $K_{1/3}(\kappa)$ and $K_{2/3}(\kappa)$ are two modified Bessel functions of the second kind, ${\rm Int}K_{1/3}(\kappa) \equiv \int_{\kappa}^{\infty}K_{1/3}(x)dx$, $\kappa=2u/[3(1-u)\chi_e]$, $u=\varepsilon_\gamma / \varepsilon_e$, $\varepsilon_\gamma$ is the photon energy, and $\varepsilon_e$ is the electron energy before photon emission.

Photon polarization is characterized by the Stokes vector $\bm \xi = (\xi_1, \xi_2, \xi_3)$ in the orthogonal basis of ($\hat{\bm e}_1$, $\hat{\bm e}_2$, $\hat{\bm e}_v$) \cite{baier1998qed,cain}, where $\hat{\bm e}_v$ is the velocity direction of the emitting electron, $\hat{\bm e}_1$ is its transverse acceleration direction, and $\hat{\bm e}_2 = \hat{\bm e}_v \times \hat{\bm e}_1$. For a linearly polarized laser interacting with the unpolarized plasma, the $\xi_1$ and $\xi_2$ components are zero on average, leaving only the $\xi_3$ component in Eq.~\eqref{eq1}—referred to as \emph{photon polarization} \cite{song2021njp,song2022prl}. The mixed polarization state $|\xi_3| \leq 1$ represents the mean polarization of an ensemble of many real photons, or a macrophoton \cite{cain}, which aligns with the PIC method. Unless otherwise specified, the term ``photon'' refers to this macrophoton. In particular, a photon with polarization $\xi_3$ means that a fraction of $(1 + \xi_3)/2$ of real photons are linearly polarized along $\hat{\bm e}_1$, while a fraction of $(1 - \xi_3)/2$ are polarized along $\hat{\bm e}_2$. Thus, the sign of $\xi_3$ determines the predominant orientation of photon polarization. In our numerical module, the photon emission direction is assumed to be collinear with the electron velocity. To simplify notation, we also use $\hat{\bm e}_v$ to represent the photon velocity.

\begin{figure}[!t]
\centering
\includegraphics[width=0.48\textwidth]{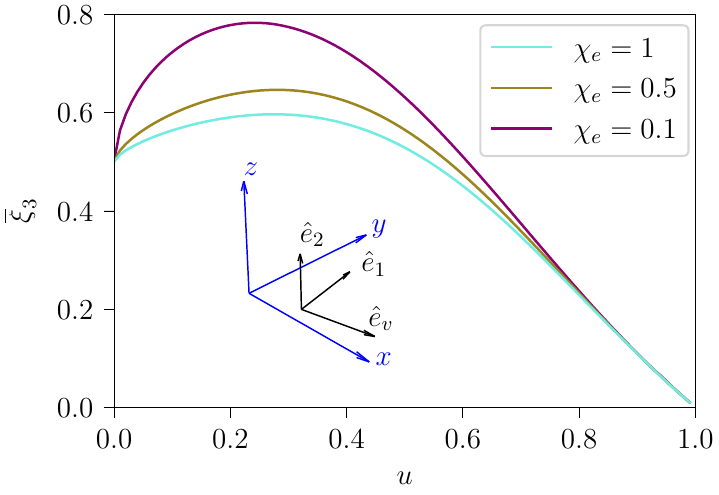}
\caption{\label{fig1} The average photon polarization $\overline\xi_3$, as a function of the photon energy ratio $u = \varepsilon_\gamma/\varepsilon_e$, is plotted according to Eq.~\eqref{eq2} for three different QED parameters, $\chi_e = 0.1$, 0.5, and 1. The inset illustrates the Stokes basis ($\hat{\bm e}_1$, $\hat{\bm e}_2$, $\hat{\bm e}_v$), with the laser field linearly polarized in the $x$-$y$ plane.}
\end{figure}

In principle, the Stokes bases for each photon are different, so they cannot be directly compared. However, for the linearly polarized laser field we focus on, the vectors $\hat{\bm e}_1$ and $\hat{\bm e}_v$ of all photons lie within the laser polarization plane, which holds strictly in the one-dimensional (1D) and two-dimensional (2D) PIC setups. Consider a laser propagating along the $+x$ direction and linearly polarized along the $y$ direction, consistent with the configuration adopted in Sec.~\ref{setup}. Under the action of the linearly polarized laser field, electrons move within the laser polarization plane, i.e., the $x$-$y$ plane. The transverse acceleration direction $\hat{\bm e}_1$ of the electrons is also parallel to the $x$-$y$ plane, so $\hat{\bm e}_2 = \hat{\bm e}_v \times \hat{\bm e}_1$ is directed along the $z$-axis, as illustrated in the inset of Fig.~\ref{fig1}. As a result, photon polarization $\xi_3$ in the linearly polarized laser field can be interpreted as follows: $(1 + \xi_3)/2$ of real photons are linearly polarized parallel to the laser polarization plane, denoted as $\gamma_\parallel$ photons, and $(1 - \xi_3)/2$ are linearly polarized perpendicular to the laser polarization plane, denoted as $\gamma_\perp$ photons.

Derived from Eq.~\eqref{eq1}, the average photon polarization $\overline\xi_3$ as a function of $\chi_e$ and $u$ is \cite{song2021njp}
\begin{eqnarray}\label{eq2}
\overline\xi_3=\frac{K_{2/3}(\kappa)}{\frac{u^2-2u+2}{1-u}K_{2/3}(\kappa)-{\rm Int}K_{1/3}(\kappa)}.
\end{eqnarray}
Equation~\eqref{eq2} is visualized in Fig.~\ref{fig1} for three different QED parameters: $\chi_e = 0.1$, 0.5, and 1. It shows that photons emitted by unpolarized electrons are highly linearly polarized, with the average photon polarization $\overline\xi_3 > 0$, indicating that photons are primarily polarized along $\hat{\bm e}_1$. As the photon energy ratio $u = \varepsilon_\gamma / \varepsilon_e$ increases, $\overline\xi_3$ rises from 50\% at $u \rightarrow 0$, reaching a maximum of 0.6–0.8 at $u \approx 0.2$–0.4, and then decreases to 0 as $u \rightarrow 1$. Additionally, $\overline\xi_3$ decreases with increasing $\chi_e$ at a given $u$, indicating that photon polarization weakens as $\chi_e$, which is mainly determined by the laser intensity, increases.

\subsection{Photon-polarization-resolved LBW pair production}
\label{LBW}

\begin{figure*}[t]
\centering
\includegraphics[width=\textwidth]{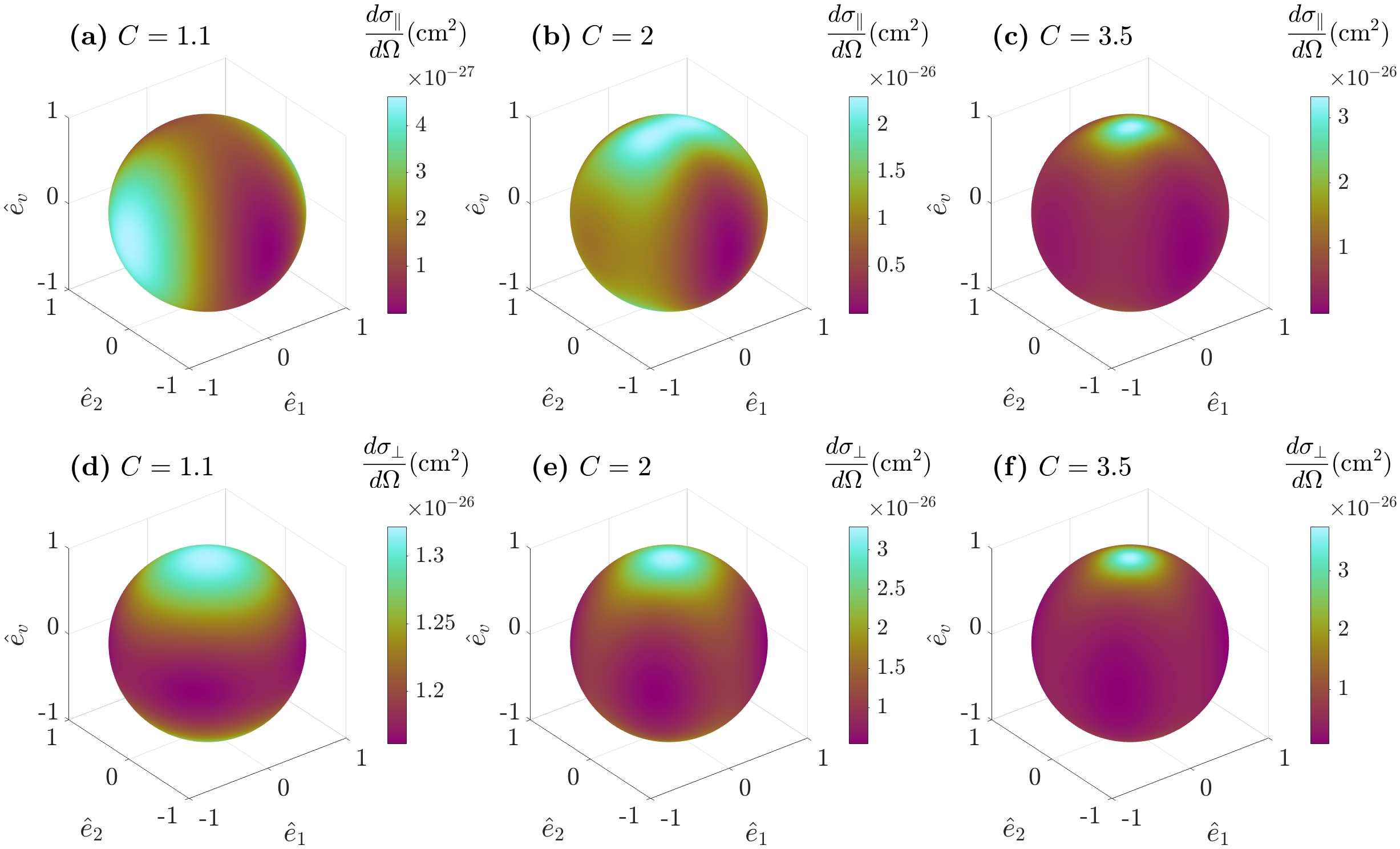}
\caption{\label{fig2} The theoretical LBW cross sections, with the positron emission angle resolved, for parallel photon polarization [(a)--(c), where both photons are linearly polarized along the $\hat{\bm e}_1$ direction] and perpendicular photon polarization [(d)--(f), with one photon linearly polarized along the $\hat{\bm e}_1$ direction and the other along the $\hat{\bm e}_2$ direction] are shown for three different CM photon energies $C = 1.1$ [(a), (d)], $C = 2$ [(b), (e)], and $C = 3.5$ [(c), (f)], as calculated using Eqs.~\eqref{eq3} and \eqref{eq4}.}
\end{figure*}

In the pioneering work \cite{breit1934pr}, Breit and Wheeler derived the LBW cross section for two types of pure linear polarization states. In the CM frame, the linear polarizations of two colliding photons are represented by $\xi_3'$ and $\xi_3''$, with the primes distinguishing the first and second photon, respectively. The emission direction of the generated positron, $\hat{\bm e}_+$, is defined in terms of a polar angle $\theta$ (relative to the $\hat{\bm e}_v$ axis) and an azimuthal angle $\varphi$ (in the plane spanned by $\hat{\bm e}_1$ and $\hat{\bm e}_2$), such that $\hat{\bm e}_+ = \cos\theta~\hat{\bm e}_v + \sin\theta\cos\varphi~\hat{\bm e}_1 + \sin\theta\sin\varphi~\hat{\bm e}_2$ and the corresponding solid angle $d\Omega=\sin\theta d\theta d\varphi$. The LBW cross sections for parallel photon polarization ($\xi_3' = 1$ and $\xi_3'' = 1$) and perpendicular photon polarization ($\xi_3' \xi_3'' = -1$) are expressed as \cite{breit1934pr}
\begin{eqnarray}\label{eq3}
\frac{d\sigma_\parallel}{d\Omega}(\theta,\varphi)&=&\sigma_0\left[S^2 - 2 S^2(N^2 - M^2)-S^4 (N^2 - M^2)^2 + S^2 C^2 (1 - \Lambda^2)\right],\\
\label{eq4}
\frac{d\sigma_\perp}{d\Omega}(\theta,\varphi)&=&\sigma_0\left[C^2 - 4 S^4 M^2 N^2+S^2 C^2 (1 - \Lambda^2)\right],
\end{eqnarray}
where $\sigma_0=r_e^2S/[2 C^3 (C^2-\Lambda^2 S^2)^2]$ and $r_e=e^2/m_ec^2\approx2.82\times10^{-13}$ cm is the classical electron radius. Here, $C\equiv\cosh\Theta$ represents the photon energy normalized to $m_ec^2$ in the CM frame---hereafter referred to as \emph{CM photon energy}, and $S\equiv\sinh\Theta$. The cosine values of the angles between the positron emission direction $\hat{\bm e}_+$ and the vectors $\hat{\bm e}_1$, $\hat{\bm e}_2$, and $\hat{\bm e}_v$ are denoted by $M$, $N$, and $\Lambda$, respectively. Therefore, the relation $M^2 + N^2 + \Lambda^2 = 1$ holds.

According to Eqs.~\eqref{eq3} and \eqref{eq4}, the LBW cross sections for two types of photon polarization at three different CM photon energies are illustrated in Figs.~\ref{fig2}(a)--\ref{fig2}(f). For the low CM photon energy of $C=1.1$, parallel photon polarization significantly affects the angular distribution of positron emission, with positrons being preferentially emitted along the photon polarization direction $\hat{\bm e}_1$ [Fig.~\ref{fig2}(a)]. In contrast, perpendicular photon polarization has a small effect on the positron emission direction for the small $C$, with positrons being emitted somewhat preferentially along the photon velocity direction [Fig.~\ref{fig2}(d)]. As the CM photon energy increases to $C=3.5$, the polarization effect weakens, and positron emission becomes more concentrated along the photon velocity direction $\hat{\bm e}_v$ for both polarization types [Figs.~\ref{fig2}(c) and \ref{fig2}(f)].

By summing over the positron emission angle $\theta$ and $\varphi$ in Eqs.~\eqref{eq3} and \eqref{eq4}, the total LBW cross sections for both parallel and perpendicular photon polarizations can be obtained \cite{breit1934pr}:
\begin{eqnarray}\label{eq5}
\sigma_\parallel&=&\frac{\pi r_e^2}{4}\left(4\Theta C^{-2}+4\Theta C^{-4}-3\Theta C^{-6}-2SC^{-3}-3SC^{-5}\right),\\
\label{eq6}
\sigma_\perp&=&\frac{\pi r_e^2}{4}\left(4\Theta C^{-2}+4\Theta C^{-4}-\Theta C^{-6}-2SC^{-3}-SC^{-5}\right).
\end{eqnarray}

In a linearly polarized laser field, Eqs. \eqref{eq3}--\eqref{eq6} are readily generalized to partially polarized photons. The collision of two partially polarized photons with polarization $\xi_3'$ and $\xi_3''$ involves four distinct classes of real photons according to whether their polarization is parallel or perpendicular to the laser polarization plane: $\gamma'_\parallel$, $\gamma'_\perp$, $\gamma''_\parallel$, and $\gamma''_\perp$. Following Lorentz transformation to the CM frame, $\gamma'_\parallel$ and $\gamma''_\parallel$ photons maintain parallel polarization \cite{cocke1972nps}, accounting for a fraction of $(1+\xi_3')(1+\xi_3'')/4$ of the total real photons. Similarly, $\gamma'_\perp$ and $\gamma''_\perp$ photons also preserve parallel polarization, contributing $(1-\xi_3')(1-\xi_3'')/4$. When one photon is polarized parallel to the laser polarization plane and the other perpendicular, such as $\gamma'_\parallel$ and $\gamma''_\perp$ photons, or $\gamma'_\perp$ and $\gamma''_\parallel$ photons, their mutual perpendicular polarization persists in the CM frame, contributing fractions $(1+\xi_3')(1-\xi_3'')/4$ and $(1-\xi_3')(1+\xi_3'')/4$, respectively. Finally, the LBW cross section, with positron emission angle and photon polarization resolved, is formulated as
\begin{eqnarray}\label{eq7}
\frac{d\sigma}{d\Omega}(\theta,\varphi)&=&\frac{1+\xi_3'}{2}\frac{1+\xi_3''}{2}\times\frac{d\sigma_{\parallel}}{d\Omega}(\theta,\varphi)+\frac{1-\xi_3'}{2}\frac{1-\xi_3''}{2}\times\frac{d\sigma_{\parallel}}{d\Omega}(\theta,\varphi+\pi/2)\nonumber\\&&+\left[\frac{1+\xi_3'}{2}\frac{1-\xi_3''}{2}+\frac{1-\xi_3'}{2}\frac{1+\xi_3''}{2}\right]\times\frac{d\sigma_{\perp}}{d\Omega}(\theta, \varphi).
\end{eqnarray}
Similarly, the photon-polarization-resolved total LBW cross section is written as
\begin{eqnarray}\label{eq8}
\sigma&=&\left(\frac{1+\xi_3'}{2}\frac{1+\xi_3''}{2}+\frac{1-\xi_3'}{2}\frac{1-\xi_3''}{2}\right)\times\sigma_\parallel+\left(\frac{1+\xi_3'}{2}\frac{1-\xi_3''}{2}+\frac{1-\xi_3'}{2}\frac{1+\xi_3''}{2}\right)\times\sigma_\perp\nonumber\\&=&\frac{\pi r_e^2}{4}\left[4\Theta C^{-2}+4\Theta C^{-4}-(2+\xi_3'\xi_3'')\Theta C^{-6}-2SC^{-3}-(2+\xi_3'\xi_3'')SC^{-5}\right].
\end{eqnarray}

Using the two basic identities $\cosh^2x-\sinh^2x=1$ and $\operatorname{arcosh} x=\ln\left(x+\sqrt{x^2-1}\right)$, the photon-polarization-averaged total LBW cross section $\overline\sigma$ can be expressed into a commonly used form \cite{jauch1976book}:
\begin{equation}\label{eq9}
\overline\sigma=\frac{\pi r_e^2}{2}(1-\nu^2)\left[2\nu^3-4\nu+(3-\nu^4)\ln\left(\frac{1+\nu}{1-\nu}\right)\right],
\end{equation}
where $\nu=\sqrt{1-1/C^2}$.

\begin{figure}[t]
\centering
\includegraphics[width=0.48\textwidth]{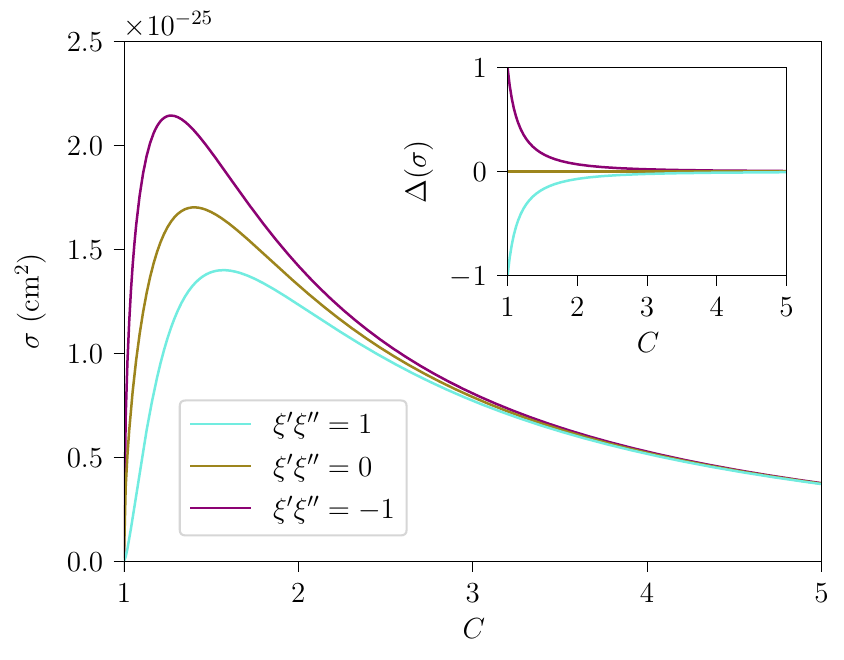}
\caption{\label{fig3} The total LBW cross section $\sigma$ as a function of the CM photon energy $C$ for parallel ($\xi_3'\xi_3''=1$), perpendicular ($\xi_3'\xi_3''=-1$), and unpolarized ($\xi_3'\xi_3''=0$) photon polarizations, plotted based on Eq.~\eqref{eq8}. The relative deviation of the polarized cross section from the unpolarized result, $\Delta(\sigma) = (\sigma^{\rm pol} - \sigma^{\rm unpol}) / \sigma^{\rm unpol}$, is shown in the inset.}
\end{figure}

Figure~\ref{fig3} shows the total LBW cross sections, calculated using Eq.~\eqref{eq8}, for parallel, perpendicular, and unpolarized photon polarizations. Parallel polarization results in a reduced cross section compared to the unpolarized case at the same CM photon energy $C$. In contrast, perpendicular polarization increases the cross section. For both parallel and perpendicular polarizations, the absolute relative variation between the polarized and unpolarized results, $\Delta(\sigma) = (\sigma^{\rm pol} - \sigma^{\rm unpol}) / \sigma^{\rm unpol}$, decreases from 100\% at $C \rightarrow 1$ to 0 as $C \gg 1$. Therefore, the polarization effect on LBW pair production is most pronounced by low-energy photons.

\subsection{PIC implementation}
\label{pic}

\begin{figure*}[t]
\centering
\includegraphics[width=\textwidth]{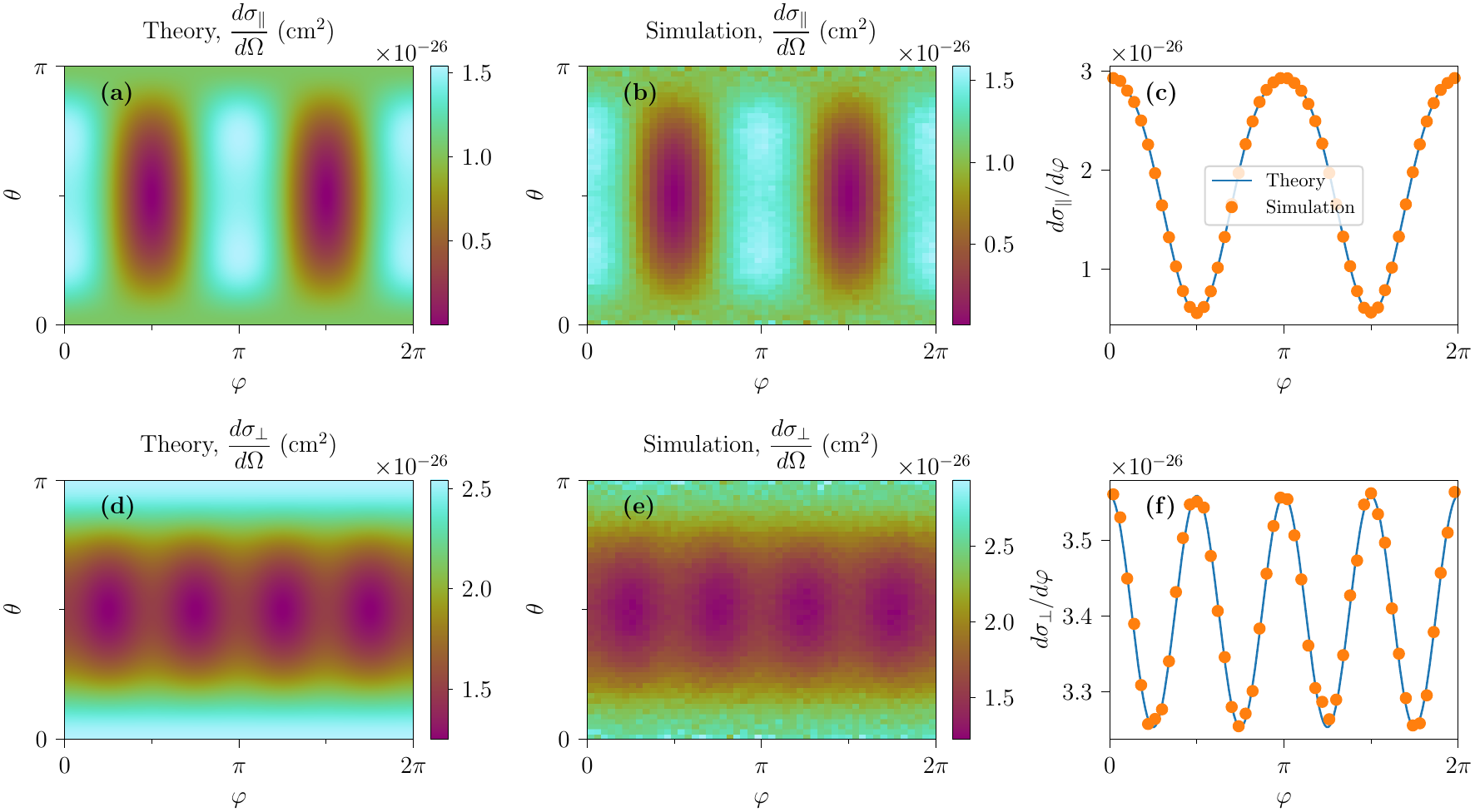}
\caption{\label{fig4} The comparison of LBW cross sections between theories and simulations for parallel [(a)--(c)] and perpendicular [(d)--(f)] photon polarizations. }
\end{figure*}

The photon-polarization-resolved nonlinear Compton scattering has been implemented into the {\scshape yunic} PIC code \cite{song2021arxiv} in our previous works \cite{song2021njp,song2022prl}. Here, we introduce how to implement photon-polarization-resolved LBW pair production into the 1D/2D PIC code using the Monte-Carlo sampling based on Eq.~\eqref{eq7}.

In the linearly polarized laser field and 1D/2D PIC setup, the motion of electrons and photons is confined to the laser polarization plane. Therefore, polarization $\xi_3$ of emitted photons calculated via Eq.~\eqref{eq1} can be directly inserted into Eq.~\eqref{eq7} without the need for additional transformation, according to Sec.~\ref{LBW}. Specifically, three uniform random numbers $U_1$, $U_2$ and $U_3$ between 0 and 1 are required to determine if an $e^\pm$ pair is generated during the photon-photon collision and its possible angular distribution. The polar and azimuthal angles of the generated positron are sampled by $\theta=U_1\times\pi$ and $\varphi=U_2 \times 2\pi$. If the pair production probability $P_{\rm LBW}=\pi\times 2\pi\times\mathbb{F}_{\rm mult}\mathbb{N}_{\rm ratio}{\rm Max}(w_1,w_2)\frac{d\sigma}{d\Omega}(\theta,\varphi)(1-\cos\theta_{\gamma\gamma})c\Delta t/\Delta V$ is greater than $U_3$, then an $e^\pm$ pair with the weight of ${\rm Min}(w_1,w_2)/\mathbb{F}_{\rm mult}$ is created; otherwise, pair production is rejected. Here, $w_{1,2}$ denotes the weight of two colliding photons, $\Delta t$ the PIC time step, $\Delta V$ the PIC cell volume, $\mathbb{N}_{\rm ratio}$ is to compensate for not making all possible pairings for efficiency, and $\mathbb{F}_{\rm mult}$ is the multiplication factor that actually does not affect the positron yield. The pairwise collision of photons with different weights was detailed in \cite{song2024pre}, including the functional roles of two factors $\mathbb{N}_{\rm ratio}$ and $\mathbb{F}_{\rm mult}$ \cite{higginson2019jcp}.

We benchmark our photon-polarization-resolved LBW algorithm through simulating the collision between two linearly polarized photon beams using 1D version of the {\scshape yunic} code. The simulation is performed in a 10 $\mu$m domain discretized into 320 uniform grid cells. Periodic boundary conditions are applied for both fields and particles. Two monoenergetic photon beams, each with an uniform density of $1.1 \times 10^{24} / \rm cm^3$ and an energy of 1.3 MeV, collide head-on. Figures~\ref{fig4}(a)--\ref{fig4}(c) and \ref{fig4}(d)--\ref{fig4}(f) show the angular number distribution of generated positrons for parallel and perpendicular photon polarizations. Quantitative comparison reveals that our simulation results demonstrate excellent agreement with theoretical predictions for both polarization types.

\section{Simulation Parameters and Results}
\label{simulation}

\subsection{Simulation setup}
\label{setup}

To investigate the effect of photon polarization on LBW pair production in the laser-plasma interaction, we perform PIC simulations using the 2D version of the {\scshape yunic} code. We model a thick, fully ionized solid target, with the target's front positioned at $x=10~\mu$m and its rear extending to the right boundary of the simulation domain. The electron density of the bulk plasma is set to $200n_c$, where $n_c = m_e \omega_0^2 / 4\pi e^2$ is the critical plasma density and $\omega_0$ is the laser angular frequency. To account for laser prepulses in real experiments, a low-density preplasma with an exponential scale length of $L_0 = 1~\mu$m is assumed at the target's front. A laser pulse, linearly polarized along the $y$ direction, is normally incident from the left boundary onto the target. The laser has a central wavelength of $\lambda_0 = 1~\mu$m, a normalized amplitude $a_0 = eE_0 / m_e c \omega_0 = 100$–$300$, a FWHM duration of $5T_0$ ($T_0 = \lambda_0 / c \approx 3.3$ fs), and a waist radius of $3\lambda_0$. The simulation domain is set to $L_x \times L_y = 15~\mu$m $\times$ 15~$\mu$m, with a resolution of $480 \times 480$ cells. The number of ions per cell is fixed at 36, while the number of electrons per cell depends on the laser intensity: 1225 for $a_0 = 100$, 900 for $a_0 = 200$, and 625 for $a_0 = 300$. The increased number of electrons per cell at lower laser intensities is aimed at improving the statistical accuracy of generated LBW pairs. Absorbing boundaries are applied for both fields and particles in each direction. Only photons with energy exceeding $0.1m_e c^2$ are saved in our simulations. Based on our recent work \cite{song2024pre} using similar parameters, the LBW process dominates over the nonlinear LBW process for normalized laser amplitudes $a_0 < 400$–500, and thus, only the LBW process is discussed here.

\subsection{Simulation results}
\label{results}

\begin{figure*}[!t]
\centering
\includegraphics[width=\textwidth]{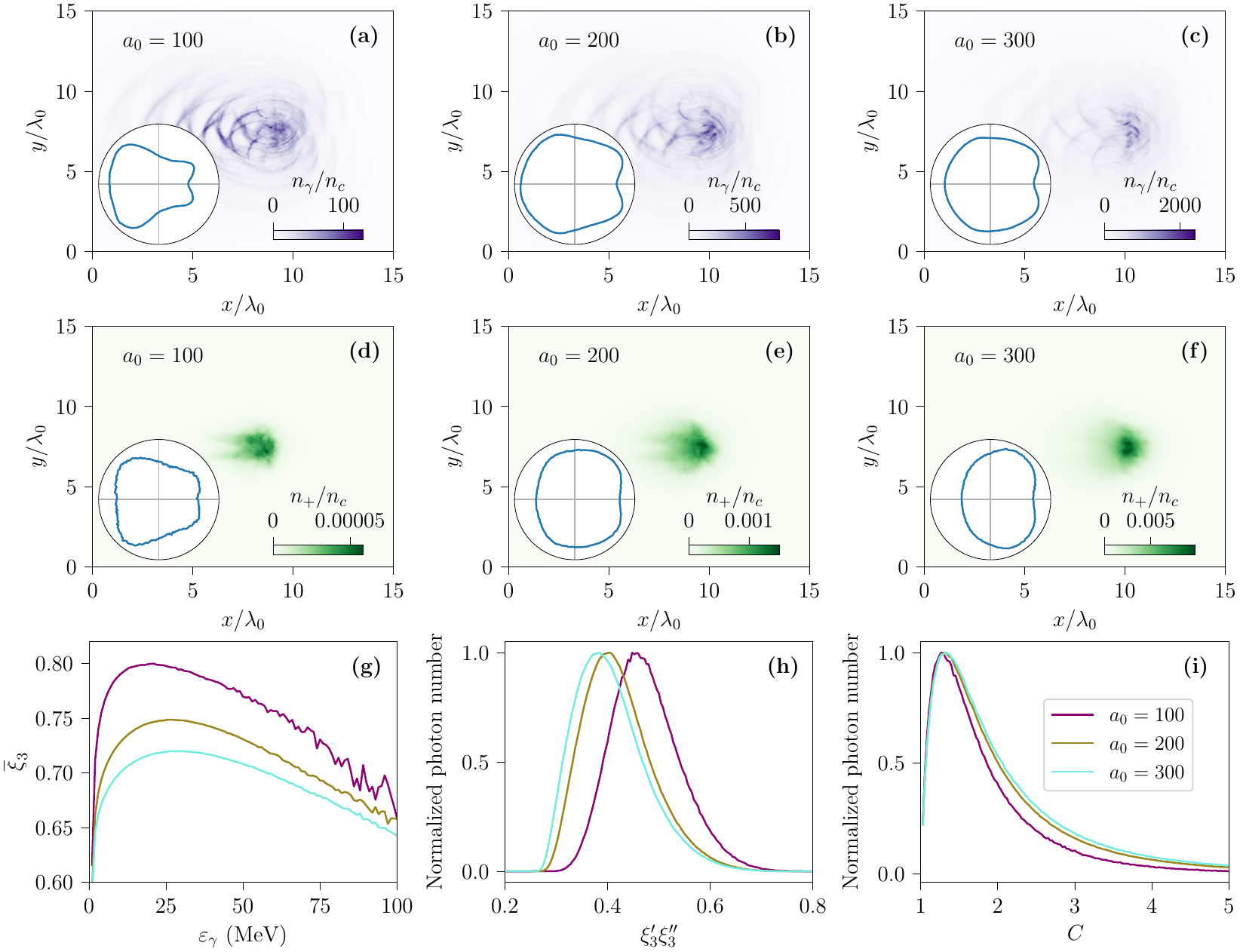}
\caption{\label{fig5} PIC simulation results. (a)--(f) Snapshots of photon density $n_\gamma$ with photon energy greater than 0.1 MeV  at $t=22T_0$ [(a)--(c)] and positron density $n_+$ at their creation time [(d)--(f)] for three different laser intensities: $a_0=100$ [(a), (d)], 200 [(b), (e)], and 300 [(c), (f)]. The inserts in (a)--(c) and (d)--(f) show the corresponding angular number distribution of photons and positrons in the $x$-$y$ plane, recorded at the end of the interaction $t=29T_0$. (g) The average photon polarization $\overline\xi_3$ as a function of the photon energy $\varepsilon_\gamma$ for three laser intensities. The normalized number distribution of annihilation photons as a function of the polarization product $\xi_3'\xi_3''$ and CM photon energy $C$ is displayed in (h) and (i), respectively.}
\end{figure*}

Figures~\ref{fig5}(a)–\ref{fig5}(f) display snapshots of photon density at $t = 22T_0$ and positron density at the time of creation for three different laser intensities. The positron density, less than $0.01n_c$ for $a_0<300$, is much lower than the target plasma density, indicating that the feedback from generated $e^\pm$ pairs on laser-plasma dynamics can be neglected. Since we focus on the influence of photon polarization on LBW pair production, the subsequent motion of positrons after their creation is not considered. The angular distribution of photon numbers at the end of the interaction is shown in the insets of Figs.~\ref{fig5}(a)–\ref{fig5}(c). Overall, photon emission in the laser-solid interaction is nearly isotropic within the laser polarization plane. While this isotropy is unfavorable for the application of $\gamma$ photons, it is highly beneficial for the LBW process, in which $\gamma$ photon collisions are essential. More specifically, photon emission shifts from being predominantly backward-directed to forward-directed as the laser intensity increases from $a_0=100$ to 300. Similarly, the angular distribution of positron emission mirrors that of $\gamma$ photons, as shown in the insets of Figs.~\ref{fig5}(d)--\ref{fig5}(f).

\begin{figure}[t]
\centering
\includegraphics[width=0.48\textwidth]{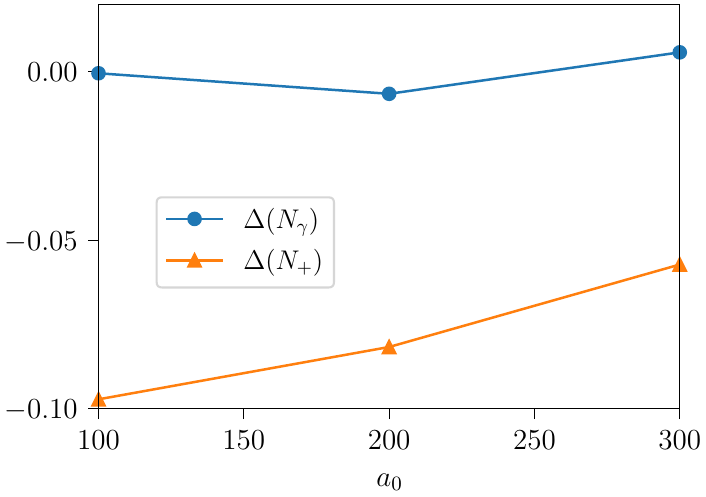}
\caption{\label{fig6} The relative variations in the total photon number and total positron number between polarized and unpolarized cases, $\Delta(N_{\gamma}) = (N_{\gamma}^{\rm pol} - N_{\gamma}^{\rm unpol}) / N_{\gamma}^{\rm unpol}$ and $\Delta(N_{+}) = (N_{+}^{\rm pol} - N_{+}^{\rm unpol}) / N_+^{\rm unpol}$, for three laser intensities.}
\end{figure}

Figure~\ref{fig5}(g) illustrates that emitted photons are highly linearly polarized. The average photon polarization $\overline\xi_3>0$ indicates that linear polarization of emitted photons is predominantly parallel to the laser polarization plane. As the photon energy increases, $\overline\xi_3$ initially increases, reaching a peak of approximately 0.8, 0.75, and 0.72 for laser intensities $a_0 = 100$, 200, and 300, respectively. As the laser intensity increases, photon polarization decreases across the entire energy range, consistent with Fig.~\ref{fig1}. The high polarization of emitted photons results in the polarization product $\xi_3' \xi_3'' > 0$ for LBW pair production, as both $\xi_3' > 0$ and $\xi_3'' > 0$ for two colliding photons. The normalized number distribution of annihilation photons as a function of the polarization product is shown in Fig.~\ref{fig5}(h). The polarization product associated with the highest number of annihilation photons is approximately 0.45, 0.4, and 0.38 for $a_0 = 100$, 200, and 300, respectively. Another important parameter for assessing the impact of photon polarization is the CM photon energy. The CM energy of annihilation photons is primarily concentrated in the low-energy range, peaking at around $C = 1.3$ for all three laser intensities, as shown in Fig.~\ref{fig5}(i).

The positive polarization product and low CM photon energy suggest that photon polarization will have a significant impact on LBW pair production according to Fig.~\ref{fig3}. When photon polarization is taken into account, our PIC simulation results indicate that the total LBW positron yield is reduced by approximately 5\%--10\% compared to the unpolarized case. With the increase of laser intensities, the absolute relative variation in the total positron yield between polarized and unpolarized cases, $\Delta(N_+)$, decreases from over 10\% at $a_0=100$ to about 6\% at $a_0=300$, as shown in Fig.~\ref{fig6}. There are two main reasons for this decrease: (i) the photon polarization product decreases [Fig.~\ref{fig5}(i)]; (ii) the CM photon energy increases [Fig.~\ref{fig5}(j)]. According to theoretical predictions shown in Fig.~\ref{fig3}, both the increase of the CM photon energy and the decrease of the polarization product reduce the influence of photon polarization on the LBW cross section, which is consequently reflected in $\Delta(N_+)$. The negligible photon number variation, $|\Delta(N_\gamma)| < 0.01$ in Fig.~\ref{fig6}, confirms that the reduction in the positron yield is due to photon polarization, rather than changes in photon numbers due to the Monte-Carlo noise.

\begin{figure*}[t]
\centering
\includegraphics[width=\textwidth]{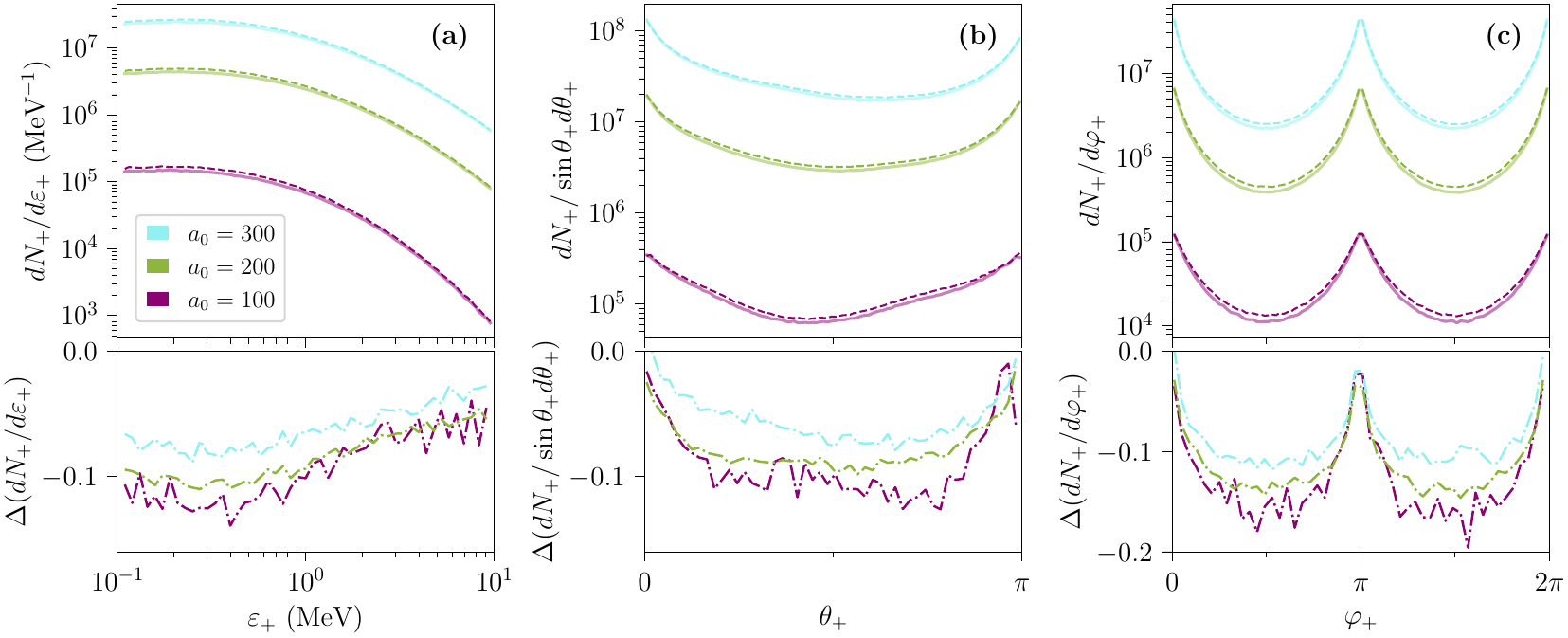}
\caption{\label{fig7} The positron number distribution as a function of the positron energy $\varepsilon_+$ (a), polar angle of positron emission $\theta_+$ relative to the $+x$ axis (b), and azimuthal angle of positron emission $\varphi_+$ measured counterclockwise from the $+y$ axis with respect to $+x$ axis (c) for three different laser intensities. The results in (a)--(c) are presented for polarized (solid) and unpolarized (dashed) cases, and their relative variations (dashed-dotted).}
\end{figure*}

The suppression of positron production by photon polarization also depends on the positron energy and the positron emission direction. Figure~\ref{fig7}(a) illustrates that the positron energy at the creation time is relatively low, as low-energy photons dominate in nonlinear Compton scattering. In agreement with Fig.~\ref{fig3}, the impact of photon polarization on low-energy positrons is generally more significant than on high-energy positrons, i.e., $|\Delta(dN_+/d\varepsilon_+)|$ decreases as the positron energy increases. The influence of photon polarization on the positron number is most pronounced at $\theta_+ = \pi/2$ and $\varphi_+=\pi\pm\pi/2$, as shown in Figs.~\ref{fig7}(b) and \ref{fig7}(c). This indicates that photon polarization primarily affects positron emission in directions outside the laser polarization plane, while its effect within the laser polarization plane is relatively weak. This behavior can be explained by the theoretical LBW cross sections shown in Figs.~\ref{fig3}(a) and \ref{fig3}(b) that positrons are predominantly emitted along the photon polarization direction for parallel photon polarization, which is parallel to the laser polarization plane. Therefore, the number of positrons emitted perpendicular to the laser polarization is significantly lower compared to the unpolarized case.

Finally, we would like to add two points. First, although our comparison shows that photon polarization reduces the LBW positron yield, this does not imply that the photon polarization effect can be directly verified in experiments. It is not impossible to ``turn off'' photon polarization in real experiments. Therefore, to experimentally verify the influence of photon polarization on the LBW process, a carefully designed experimental setup is required. Second, the angle-dependent impact of photon polarization on positron production is observed at the moment of their creation. Once the subsequent influence of laser fields is considered, positrons will acquire significant momentum parallel to the laser polarization plane via vacuum laser acceleration \cite{song2025arxiv}, which will substantially disrupt this angular dependence. Therefore, in future laser-plasma experiments aimed at verifying the angle-dependent photon polarization effect on the LBW process, it will be crucial to minimize the influence of laser fields on generated positrons.

\section{conclusion}
\label{conclusion}

We have investigated the impact of photon polarization on LBW pair production during the interaction of a linearly polarized laser with solid-density plasmas. The $\gamma$
photons emitted via nonlinear Compton scattering are predominantly linearly polarized parallel to the laser polarization plane, with a polarization degree exceeding 50\%. During the collision of such polarized photons, parallel photon polarization dominates, resulting in a reduced LBW cross section relative to the unpolarized case. Our 2D PIC simulations reveal that photon polarization reduces the total LBW positron yield by 5\%–10\%. Additionally, increasing laser intensities reduces the suppression effect of photon polarization. Although we focus on the interaction between a laser and a solid-density plasma, the findings have implications for other laser-plasma systems as photons emitted through nonlinear Compton scattering are intrinsically linearly polarized.

Remarkably, when these results are combined with our previous investigations on NBW pair production \cite{song2021njp, song2022prl}, it becomes clear that photon polarization suppresses both linear and NBW processes by a similar extent (approximately 5\%–10\%) in linearly polarized laser-driven plasmas. Moreover, in both processes, the photon polarization effect weakens as the laser intensity increases.

\begin{acknowledgments}

This work was supported by the National Science Foundation of China (Grants No.~12405285, 12135009 and 11991074) and the China Postdoctoral Science Foundation (Grants No.~2023M742294). The simulations were performed on the $\pi$ 2.0 supercomputer at Shanghai Jiao Tong University.

\end{acknowledgments}

\bibliography{reference}

\end{document}